\documentclass[final,1p,times]{elsarticle}
\usepackage[utf8]{inputenc}
\usepackage{amsmath}
\usepackage{amsfonts}
\usepackage{amssymb}
\usepackage{multirow}
\usepackage{graphicx}
\usepackage{tabularx}
\usepackage{array}  
\usepackage{enumitem}
\usepackage{hhline}
\usepackage{xcolor}
\usepackage[hyphens]{url}
\usepackage{hyperref}

\usepackage[colorinlistoftodos]{todonotes}

\begin{document}

\begin{frontmatter}

\title{SmartProduct: a prototype platform for product monitoring-as-a-service, leveraging IoT technologies and the EPCIS standard}

\author[inst1]{Petros~Zervoudakis}
\ead{zervoudak@ics.forth.gr}

\author[inst1]{Maria~Plevraki}
\ead{plevraki@ics.forth.gr}

\author[inst1,inst2]{Eleftheria~Plevridi}
\ead{eleftheria@ics.forth.gr}

\author[inst1]{Alexandros~Fragkiadakis}
\ead{alfrag@ics.forth.gr}

\affiliation[inst1]{organization={Institute of Computer Science, Foundation for Research and Technology-Hellas (FORTH)},
           city={Heraklion},
           state={Crete},
           country={Greece}}

\affiliation[inst2]{organization={Department of Computer Science, University of Crete},
           city={Heraklion},
           state={Crete},
           country={Greece}}

\begin{abstract}
Internet of Things (IoT) technologies have received significant attention in recent years by encompassing a set of technologies that enable a variety of heterogeneous physical objects, called things, to interact and communicate through efficient networking protocols. These technologies have already been used in several domains such as in manufacturing, healthcare, agriculture, etc. Another domain in which IoT can be applicable and useful, is that of supply chain tracing, where products are monitored throughout the whole supply chain. IoT data collection can enable the proliferation of applications, which are able to track environmental-related information per product (e.g. storage conditions), and combine them together with traceability data in order to provide full product monitoring services.  

Traditional supply chain tracing methods (e.g. product tracing, storage conditions' monitoring, transport vehicles used, etc.) involve costly and error-prone procedures, as human involvement is often required. Severe fragmentation is also possible, as the various stakeholders involved (producers, distributors, retailers), do not use interoperable technologies and standards.

To overcome these limitations, we propose a flexible and secure prototype platform that leverages IoT technologies, jointly with the Electronic Product Code Information Service (EPCIS) standard. The IoT software/hardware modules of the platform are used to collect IoT data (i.e. geographical location, ambient temperature, humidity, etc.), while the EPCIS-based software collects and records conventional supply chain traceability data (e.g. type of product manufactured, product loaded/unloaded into/from a truck, product transformed to another product, etc.). A rich RESTful API with strong authentication/authorisation mechanisms is used to offer Product-Monitoring-as-a-Service (PMaaS) to third-party applications.

\end{abstract}

\end{frontmatter}

\section{Introduction}
In the last decade, the proliferation of IoT technologies has boosted significantly the development of modern applications used in a wide range of domains such as in manufacturing, healthcare, agriculture, smart cities, and so on. Moreover, the IoT technologies including radio frequency identification (RFID), sensors, global positioning system (GPS)~\cite{inbook}, provide a promising opportunity to build powerful systems and applications to transform the operations of many existing industrial procedures~\cite{Xu2014InternetOT}. In supply chain traceability, for example, once an RFID tag is placed on a product, suitable data can be collected, enabling the transparency of the crucial knowledge that describes its origin, the location where it was subjected to a business process, etc. 

To this end, the global EPCIS standard\footnote{\url{https://www.gs1.org/standards/epcis}} defines a scheme for unambiguous encoding of certain physical objects that enables the way to capture and share crucial stages of their lifecycle within the supply chain, providing a holistic view, both within and across enterprises. Identification of physical objects is known as one objective of IoT, and examples of such objects include trade items or logistic units~\cite{Shahid2017InternetOT}. Furthermore, physical objects' quality supervision can increase supply chain visibility using the corresponding IoT data collected from the relevant manufacturing steps~\cite{ALFIAN2020107016}. 

At this point, we provide the following definitions: (i) \emph{supply chain traceability} (SCT), the set of processes used to trace a product within the supply chain using a popular standard such as the EPCIS one, (ii) \emph{IoT monitoring}, the set of processes used for sensory data collection (e.g. GPS data, ambient temperature, vehicle acceleration, etc.), and (iii) \emph{product monitoring} (PrM), the set of processes that combine SCT data and \emph{IoT monitoring} data, in order to provide a complete view, offered as-a-service (PMaaS), of a product within the supply chain. As an example, PrM could provide the following information: ProductA was produced in LocationA, then it was stored in Silo A under conditions: [Mean temperature: 15 degrees Celsius, humidity: 35\%], then  loaded to TruckA and transported under conditions: [Mean temperature: 18 degrees Celsius, humidity: 55\%], etc (see more details in Section~\ref{sec:interface)}).

Regarding related contributions, in~\cite{fragSCM}, we described an early version of our work for product monitoring, considering IoT technologies and the EPCIS standard.
Other related contributions mainly focus on SCT (e.g.~\cite{BYUN201735, bruno}), archiving a complete traceability in a supply chain network, while, in this work, we propose a flexible and secure platform that makes feasible PMaaS using off-the-shelf IoT devices. IoT sensory data (e.g. temperature, humidity, etc.) are collected jointly with supply chain related data (e.g. product creation, product storage in a specific warehouse, product packaging, etc.).

In~\cite{BYUN201882}, the authors propose a document-based IoT platform that uses a federation of software components, complying with the EPCIS standard, for a wide range of applications, and demonstrate that the query response time performs better compared to other similar solutions, such as Fosstrack EPCIS~\cite{Floerkemeier2007RFIDAD}, and Cassandra-based EPCIS~\cite{Le2014EPCIS}. However, the processes followed within a supply chain and any potential failures (e.g. due to storage conditions) cannot be detected, as IoT data are not collected.


Meanwhile, other works (e.g. \cite{Teucke2018SharingSB}, \cite{Han2018GS1CC}, \cite{Pham2017GS1GS}) have leveraged a flexible extension mechanism, provided by the EPCIS standard that enables developers to create enriched events, in which, additional or adjusted data members can be attached to. Consequently, additional information coming from IoT data sources can be easily stored within EPCIS events, such as temperature measurements collected within a cooling supply chain industry~\cite{Teucke2018SharingSB}, IoT data (e.g. healthcare-related, AI speakers) within a vehicle connected ecosystem~\cite{Han2018GS1CC}, and data from smart parking-related sensors (e.g. active infrared sensors, RFID, etc.)~\cite{Pham2017GS1GS}. Our work differs, as we use an independent schema for the IoT related data, which allows the continuous maintenance of data repositories, and convenient upgrading to potential future versions.

Other works (e.g.~\cite{Lin_19, Zhang_20}) use blockchain-based technologies and smart contracts to store product related information gathered from the various supply chain stages; however, they focus more on the business processes related to SCT, and show little or no interest in the joint collection of IoT data.

The remainder of this paper is organized as follows. A reference scenario is described in Section~\ref{sec:ref_sc}, and the proposed platform architecture is presented in Section~\ref{sec:arch}. In Section~\ref{sec:device_arch}, we describe the IoT devices used and their interaction with the platform. The description of two end-user interfaces is provided in Section~\ref{sec:user_int}.  Finally, conclusions and further work appear in Section~\ref{sec:concl}.

\section{Reference Scenario}
\label{sec:ref_sc}
\begin{figure}
\includegraphics[scale = 0.9]{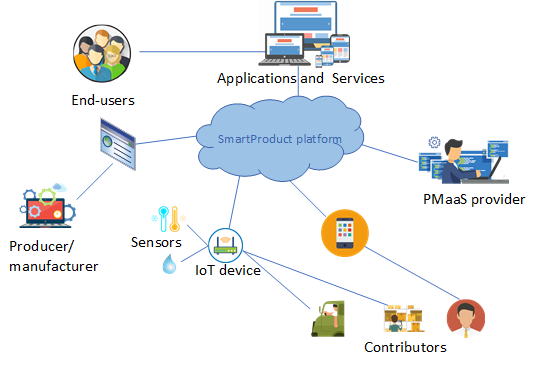}
\centering
\caption{PMaaS reference scenario}
\label{fig:reference_scenario}
\end{figure}

In Figure \ref{fig:reference_scenario}, we show the graphical representation of a scenario under consideration, as a motivation for the design and development of the \emph{SmartProduct} platform. In this scenario, a service-oriented approach is implemented for the provision and management of on-demand sensing resources, exploited to support PMaaS. Moreover, an important requirement is to offer a secure web-based user registration and authentication interface for the end-users, who can be categorised as follows: (i) PMaaS provider, who is responsible for the overall management in order to ensure platform’s proper operation, (ii) Producers/manufacturers who, upon successful registration, are able to perform specific operations such as product registration, monitoring parameters' specification, etc., (iii) End-users who are able to request specific traceability and authentication data for certain products, and (iv) Contributors, who are the IoT devices owners. Examples of contributors are individuals (or companies) who own warehouses or transport vehicles with in-premise IoT devices installed, and individuals (e.g. courier staff) who use smartphone devices to capture traceability data using an Android-based application we have developed for this purpose (presented in Section~\ref{sec:app}). Furthermore, the \emph{SmartProduct} platform can provide PMaaS to third-party applications; for example, a smart farming system that monitors the cultivation of fruits, and as soon as harvesting is completed, \emph{SmartProduct} can offer product monitoring services till the specific fruits are sold in a retail store.

We assume the existence of IoT devices, which are properly configured for secure communication, enabling product monitoring that involves IoT sensors used to measure various environmental conditions. The proposed platform collects environmental data (e.g. ambient temperature, humidity, vehicle acceleration, etc.), and GPS information, and correlates them intelligently to provide a holistic view to end-users.

\section{Product-Monitoring-as-a-service}
\label{sec:arch}
The \emph{SmartProduct} platform offers flexible and reliable PMaaS, utilizing services on a cloud infrastructure that are exposed via a rich REST API, and can also be used by authorized third-party applications. For the platform design, we considered the following requirements: (i) trustworthiness, integrity, authentication, authorization and access control, (ii) reliability, and (iii) robustness.

\subsection{Platform's main components} 
\emph{SmartProduct} is based on Openstack\footnote{\url{https://www.openstack.org}}, an open source cloud software architecture that offers a set of tools for the creation and management of an infrastructure-as-a-service, and provides a modular architecture, which gives flexibility for cloud platform's design, including integration with third-party applications.

\begin{figure*}
\includegraphics[scale = 0.58]{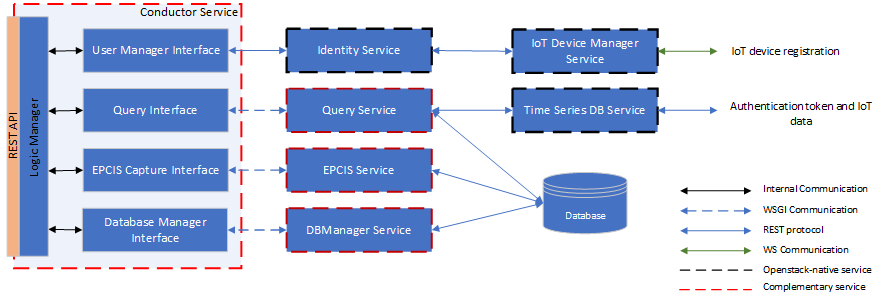}
\centering
\caption{\emph{SmartProduct} platform main components}
\label{fig:platform_architecture}
\end{figure*}

The platform (its main components are shown in Figure~\ref{fig:platform_architecture}) is designed to handle three types of events/queries: (i) \emph{EPCIS-related} events, (ii) \emph{IoT-related} events, and (iii) \emph{aggregation-related} queries. The \emph{EPCIS-related} events, e.g. sent from an EPCIS capture application such as the Android-based application presented in Section~\ref{sec:app}, convey knowledge of what happened during a step within a business process, in which physical objects (products) were handled, and are expressed via the four dimensions of what, where, when and why, as proposed by the EPCIS specification. Such events include, for example, pallet shipping, aggregation of a number of cases into a single pallet, removing cases from a pallet, etc., and business processes in which objects are fully or partially consumed and output objects are produced (e.g product transformation from bulk oil to bottled oil). The \emph{IoT-related} events include information collected by the IoT devices such as ambient temperature, humidity, vehicle velocity, geolocation information if the device is placed on a truck, etc. The \emph{aggregation-related} events are initiated by the platform users (i.e. registered producers who use platform's web-based interface, end users who use the Android application, authorized third-party applications). The platform response sent back to users, are records that contain \textit{aggregated knowledge} by combining IoT data and EPCIS-related data. For example, a product owner is able to trace his products by receiving information such as the warehouse they are stored, the trucks used to transport them, as well as the conditions during their storage and transportation, thus, providing a holistic view of the supply chain processes.   

\emph{SmartProduct} uses two types of services: (i) \emph{openstack-native} services, which are built-in Openstack services for the provision of APIs and facilitation of access resources (e.g. \emph{Identity Service}, \emph{Device Manager}, \emph{Time Series DB Service}, etc.), and (ii) \emph{complementary services}, developed by us, which support platform's significant operations (e.g. \emph{Conductor Service}, \emph{Query Service}, \emph{EPCIS Service}, \emph{DBManager Service}, etc.). Next, we provide a description of these services.

\subsubsection{Conductor Service} this \emph{complementary} service contains all platform's interfaces, through which the outside world interacts with the platform, offering the REST API endpoints, handling the requests sent by the clients, and orchestrating the access to the platform's rest of services. Each such service is accessible by a dedicated interface, allowing third-party applications to access the internal resources provided by the specific service (e.g. the \emph{EPCIS Service} is accessible through the \emph{EPCIS Capture Interface}).

\subsubsection{Identity Service}
this \emph{openstack-native} service performs centralized identity and access management based on Keystone\footnote{\url{https://docs.openstack.org/keystone/latest}}, Openstack core security framework, acting as a main authentication service across the platform. It supports a common authentication mechanism using username and password credentials, as well as token-based authorization access control (it establishes permissions based on groups and roles, allowing only authorized users to access the resources enclosed across the platform's services). After successful authentication, a token is created and allocated to a user, which is used in future service requests.

\begin{figure}[h]
\includegraphics[scale = 0.8]{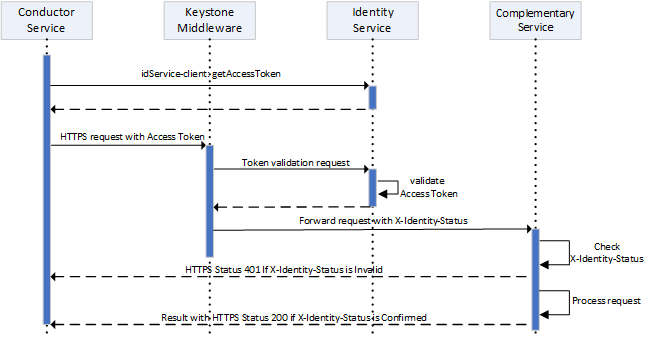}
\centering
\caption{Access control to \emph{complementary services}}
\label{fig:auth_mechanism}
\end{figure}

\subsubsection{EPCIS Service}
this is deployed as a \emph{complementary service}, and is accessible through the \emph{EPCIS Capture Interface}, which defines how EPCIS events are delivered to the EPCIS DB (Database). These events can be triggered from capture applications, such as a smartphone application. The capture interface provides a single method, which takes a single argument list, which shall contain valid EPCIS events (otherwise, they are rejected), according to this standard. This list can contain a single or multiple EPCIS events, serialized through XML binding. All relevant information such as the event time and location, Electronic Product Codes (EPCs), etc., are contained within each event. Finally, the events are stored in the EPCIS DB, a MongoDB-based implementation (which is a general-purpose, document-based, and distributed database), and become available for future use by other services. 

\subsubsection{Query Service}
this is deployed as a \emph{complementary service}, and is accessible through the \emph{Query Interface}. As mentioned previously, the \emph{SmartProduct} platform collects two types of data: (i) traceability data collected by the \emph{EPCIS Service}, following the EPCIS standard, and (ii) IoT (sensory) data, collected through the \emph{Device Manager}, which registers and authenticates the IoT devices. The main task of the \emph{Query Service} is to retrieve and combine these two different types of data, using suitable algorithms designed for this purpose. A detailed description of the algorithms used, is out of the scope of this paper.

\subsubsection{DBManager Service}
this is also deployed as a \emph{complementary service}, being accessible through \emph{the Database Manager Interface}, enabling users to register and manage their products. 
More specifically, the registration of a product can be performed at: (a) \emph{instance-level}, referring to physical products, which are handled by a process that involves \emph{instance-level} identification, where products are characterized by a unique serial identifier for that particular product (e.g. Global Trade Item (GTIN) with a serial number), and (b) \emph{batch-level}, referring to a group of products, which have a common identifier (e.g, product items packed using a single LOT number). Another functionality, provided by this service, is \emph{product discovery} that provides detailed information about the products managed by a user (e.g. list of registered products, view and edit of tracking and monitoring parameters, etc.).

\subsubsection{IoT Device Manager Service} this \emph{openstack-native} service is based on \emph{Stack4Things IoTronic}\cite{Longo201753}, an Openstack-based module, which gives the ability to register, manage, and monitor IoT devices remotely. These devices communicate with the platform through a secure TLS Websocket connection, and are authenticated and authorized through the \emph{Identity Service}, in order to appear as trusted entities within the platform. More details related to this service, and the interaction with the IoT devices, are presented in Section~\ref{sec:device_arch}.

\subsubsection{Time Series DB Service}
\label{timeseriesdb}
this is a \emph{openstack-native} service based on Openstack’s \emph{Gnocchi}\footnote{\url{https://gnocchi.osci.io}}, qualified for time series data, which are composed of event measurements, tracked, monitored and assembled over even intervals in time, and ordered sequentially. \emph{Gnocchi} is an open-source time series database, used in the \emph{SmartProduct} platform for persistent storage and indexing of time series IoT data at a large scale. 
In order to reduce the storage volume, \emph{Gnocchi} does not store the raw measurements received; rather it stores the values at specific time intervals, after applying an aggregation method as a statistical function on the raw measurements, for example, the \emph{avg} aggregation method aggregates the values of different measurements to the average value of all measurements in the time range. The way metrics are aggregated is configurable under a specific archive policy, which defines the granularity (e.g. the smallest possible time interval between two timestamps that can be stored), how long aggregates will kept in a metric, and how they will be aggregated. 

\subsection{Access control to complementary services}
As soon as a user is authenticated and a token is created, the \emph{Identity Service} enforces access control to the \emph{complementary services}. Towards this direction, the Keystone Middleware\footnote{\url{https://github.com/openstack/keystonemiddleware}} (\emph{KeM}) is used as an authentication mediator, by intersecting the incoming requests, and providing authorization capabilities through a suitable authentication protocol. 

Each \emph{complementary service} uses \emph{KeM} as the authorization mean, allowing external users to initiate requests towards it. A complete interaction for gaining access to a \emph{complementary service} is shown in Figure~\ref{fig:auth_mechanism}. \emph{KeM} is a proxy that intercepts incoming HTTPs requests and populates HTTPs headers in the request context, which are intended for the \emph{complementary services}. The token from an incoming HTTPs request is collected and validated by the Identity Service, and then \emph{KeM} populates an additional header: the HTTPs request is extended with the header X-Identity-Status, populated with the string either Confirmed or Invalid. Requests are forwarded to the \emph{complementary services} with an identity status message that indicates whether the client's identity has been confirmed or not. If confirmed, the request proceeds to further processing (for example, the \emph{DB Manager Service} is able to create, delete and edit products or the \emph{EPCIS Service} is able to store EPCIS Events).

\section{IoT devices and their interaction with the SmartProduct platform}
\label{sec:device_arch}
\begin{figure}[h]
\includegraphics[scale = 0.5]{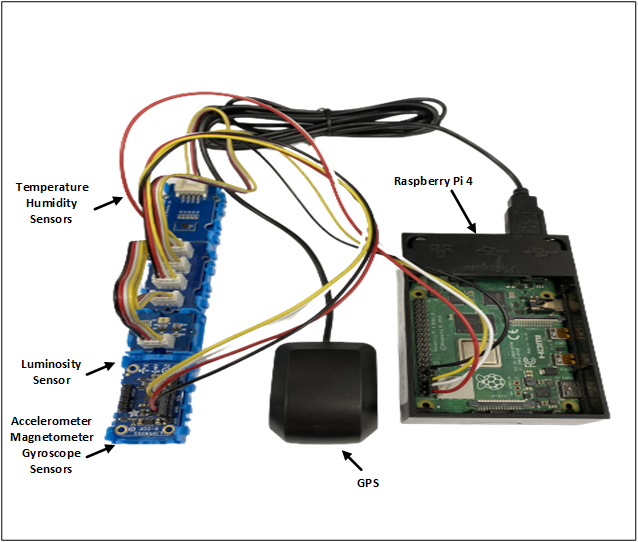}
\centering
\caption{IoT device prototype based on Raspberry Pi4, equipped with GPS and environmental sensors}
\label{fig:iot_device}
\end{figure}

In this section, we present the IoT device architecture, and we provide a detailed description of the interaction with \emph{SmartProduct}'s core services, focusing on both device and cloud components. A fleet of such IoT devices can be deployed both in fixed locations (e.g. warehouses) and in vehicles (e.g. trucks). Their prototype is based on Raspberry Pi 4\footnote{\url{https://www.raspberrypi.com}}, and they are equipped with various sensors such as GPS, temperature, humidity, ambient light, and accelerometer (Figure~\ref{fig:iot_device}). 

\begin{figure}[h]
\includegraphics[scale = 1]{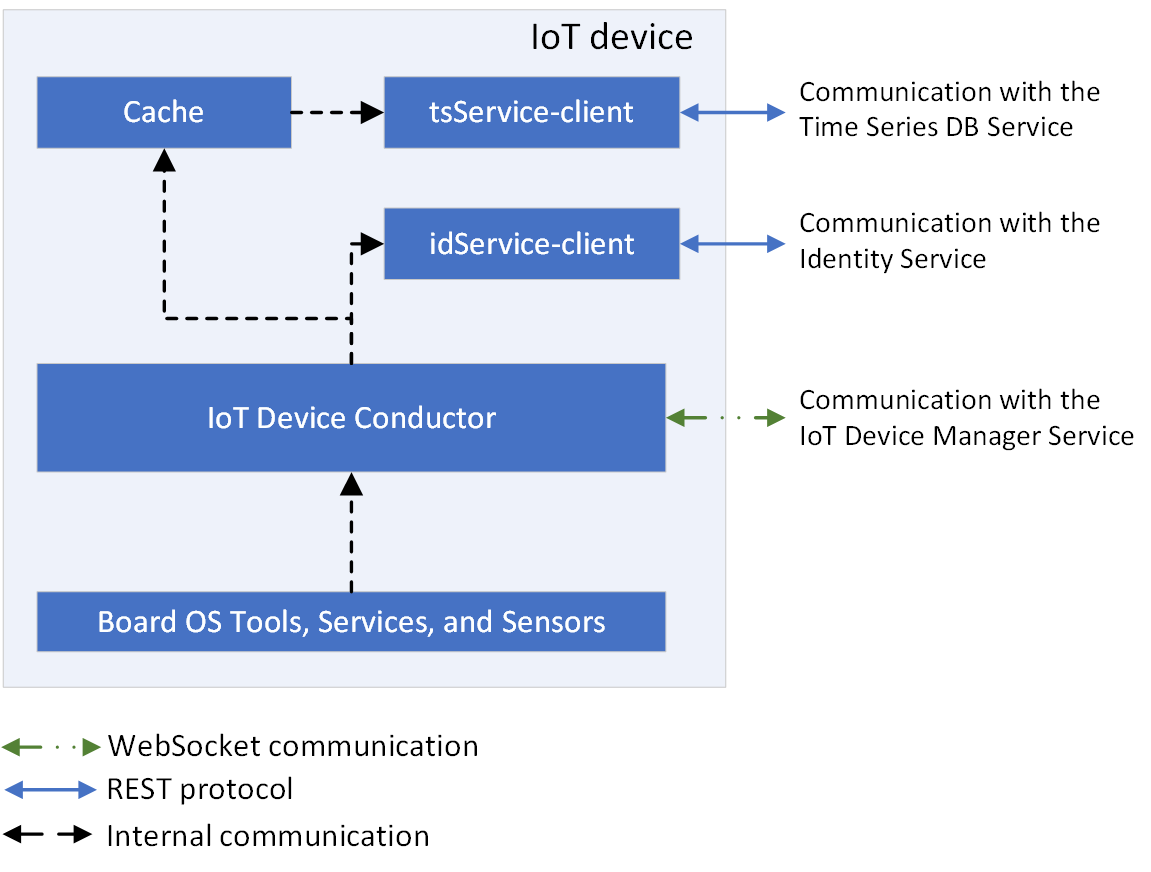}
\centering
\caption{Functional components of the IoT device}
\label{fig:device_architecture}
\end{figure}

\begin{figure}[h]
\includegraphics[scale = 0.8]{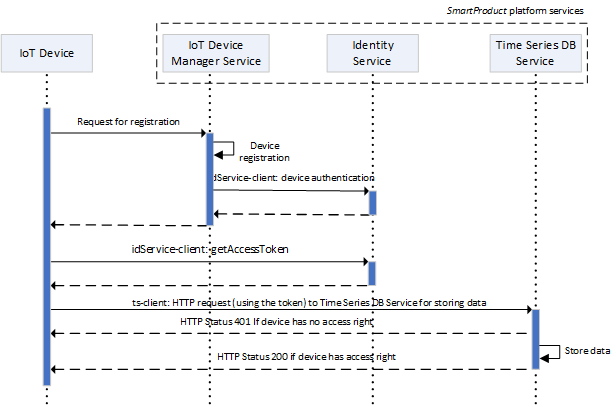}
\centering
\caption{IoT device interaction with the \emph{SmartProduct} platform}
\label{fig:seq_to_gnocchi}
\end{figure}

The main functional components of the IoT device are shown in Figure~\ref{fig:device_architecture}. The core service is the \emph{IoT Device Conductor}, which is based on \emph{IoTronic Lighting-rood}\footnote{\url{https://opendev.org/x/iotronic-lightning-rod}}, a node-side agent for the \emph{IoTronic} service. The role of this component is to orchestrate and facilitate the interaction with the \emph{SmartProduct} platform, as well as to interact with the OS tools (Raspbian), and the drivers of the sensors in order to collect the IoT data.
The complete interaction with the platform is composed of three phases (see Figure~\ref{fig:seq_to_gnocchi}):
\begin{itemize}
    \item A request is sent from the \emph{IoT Device Conductor}, through a full-duplex WebSocket channel, for the device registration. We assume that an offline process has taken place in advance, where a system administrator creates the credentials for the device in the \emph{IoTronic} service, which later on, permits its registration to the platform.
    \item A request is sent from the \emph{idService-client} component of the IoT device to the \emph{Identity Service} for device authentication, in order to appear as a trusted and authorized entity to platform's \emph{openstack-native} services. Then, the \emph{Identity Service} provides the appropriate credentials to complete the authentication process.
    \item If authentication is successful, the \emph{tsService-client} component forwards the collected IoT data to the \emph{Time Series DB Service} of the platform for persistent storage.
    
\end{itemize}

The \emph{Cache} component is used to temporarily store IoT data in case of disconnections; this enables the device to act autonomously with respect to the data collection process.
Regarding security aspects, the IoT device communicates with the platform through TLS connections with 256-bit long encryption keys.

\section{SmartProduct end-user interfaces}
\label{sec:user_int}
In this section, we present two end-user interfaces: (i) an Android-based application for recording events according to the EPCIS standard, and (ii) a web-based frontend, through which, authenticated users perform several operations regarding product registration and tracking.
\subsection{Android-based mobile application for supply chain traceability}
\label{sec:app}

\begin{figure}[h]
\includegraphics[scale = 1]{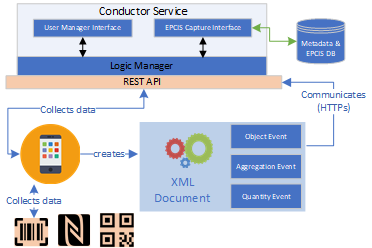}
\centering
\caption{Main functionalities of the Android-based mobile application}
\label{fig:epcis_capture_app_flow}
\end{figure}

The purpose of the mobile application is to enable SCT, and its main functionalities are (see Figure~\ref{fig:epcis_capture_app_flow}): (i) data collection from NFC, QR code, or Barcode (BR) tags that are placed on products, and (ii) communication with the \emph{SmartProduct} platform to authenticate using a cryptographic certificate and user/password credentials, in order to retrieve stored data, such as user role (i.e. producer, manufacturer), and other required information.


One of the most important features for the success of a mobile application is its User Interface (UI), which can significantly affect user experience. For the \emph{SmartProduct} platform, this mobile application has a significant role when it comes to data collection. Individuals can use the application to record SCT-related data, such as storage of products in a warehouse or transportation of pallets to another place, in order to track these products throughout the supply chain. All services are offered through a user-friendly UI. 



 The application acts as a capture medium between the information that is included within an EPC and its different interpretations, and \emph{SmartProduct} REST API, through which, this information, along with various metadata, are stored in the platform. EPC is, in essence, a numbering scheme that uniquely matches a physical object to a specific ID. There are many coding schemes, which are distinguished through their namespace in the URI prefix representation. The most common way of representing a physical object within an information system, is the Pure identity URI, which is a serial number with the form: "urn:epc:id:scheme:component1.component2; for example, “urn:epc:id:sscc” is for identifying inner packs, cases or pallets, while “urn:epc:id:sgtin” is for the product identification. EPCs can be stored in NFC tags or can be encoded with QR codes or BRs. 
 
 The selection of a specific tag type can depend on several factors such as cost, capacity, security, etc. There is always a trade-off between security and cost; for example, BR code use is much more cost efficient than NFC tag use; however, it provides no protection against product counterfeiting. Our mobile application is capable of reading EPCs from both NFC and QR/BR encoded tags. For state-of-the-art security and privacy protection, we support the use of NXP NTAG 424DNA tags\footnote{\url{https://www.nxp.com/docs/en/data-sheet/NT4H2421Gx.pdf}}.

\begin{figure}[h]
\includegraphics[scale = 0.17]{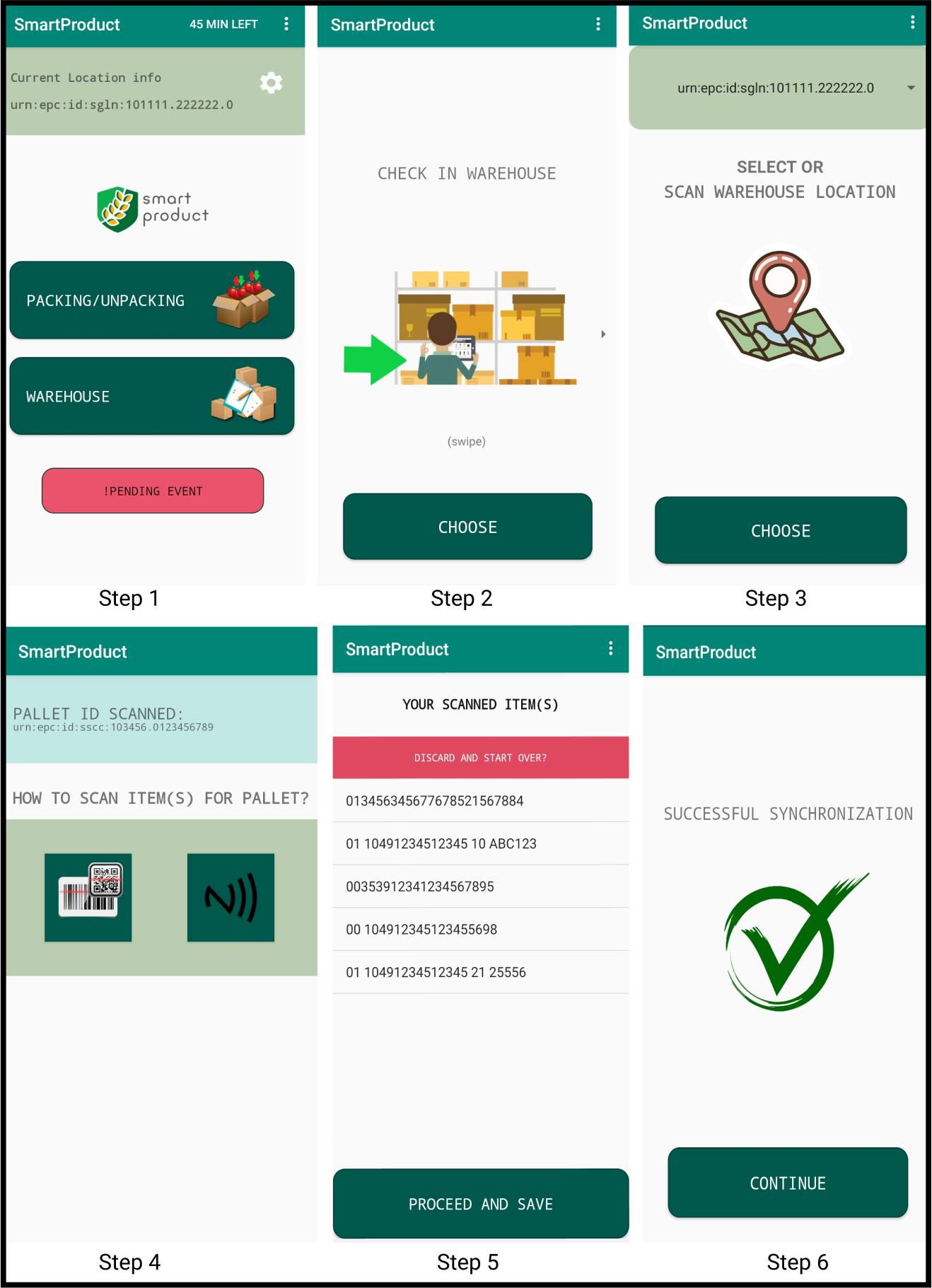}
\centering
\caption{Mobile application's UI. Steps for tracking products' storing in a warehouse}
\label{fig:application_overview}
\end{figure}

\begin{figure}[h]
\includegraphics[scale = 0.5]{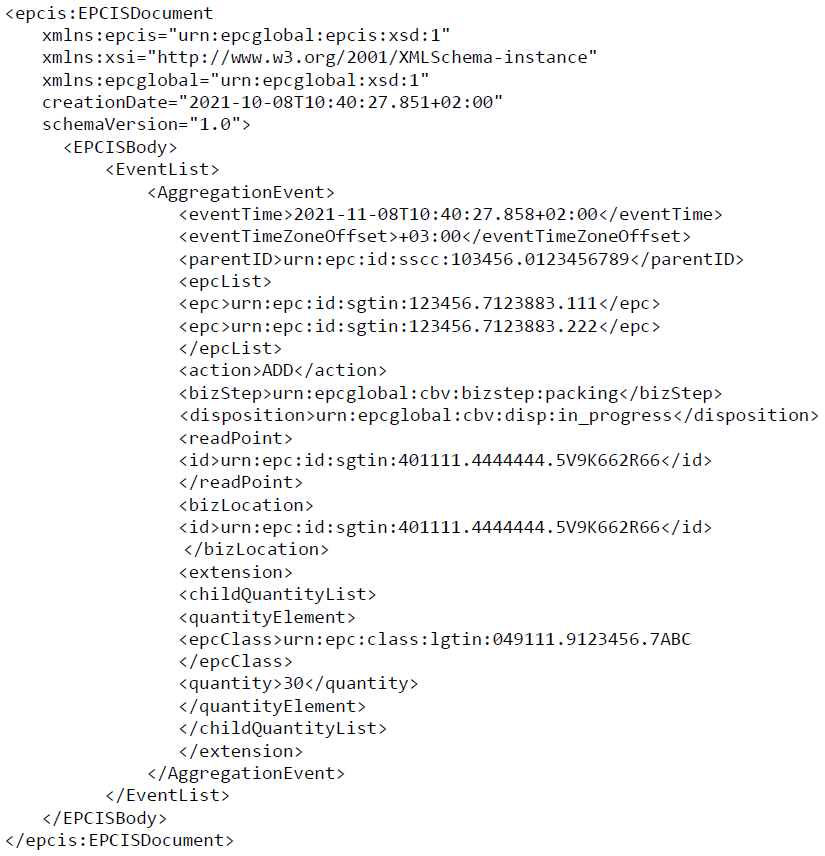}
\centering
\caption{XML document representing an EPCIS \emph{Aggregation event}}
\label{fig:epcis_aggregation_doc}
\end{figure}

In general, the EPCIS standard supports three main types of events: (i) \emph{Object event}, 
(ii) \emph{Aggregation event}, and (iii) \emph{Quantity event}. An event in the context of the EPCIS standard, has one of the previously referred types, and is represented by an XML document with a specific format (an example appears in Figure~\ref{fig:epcis_aggregation_doc}). The data included in such a file, are mainly related to EPC capturing, and contain important information about a product, such as creation/expiration date, batch or serial number, etc. An internal mapping of the EPC representation to its pure identity URI form is made for interpretation purposes. The aforementioned is partly achieved  by using Fosstrack\footnote{\url{https://fosstrak.github.io}}, an open source library for Tag Data Translation. Other data sources are the user's specific information that are retrieved from the \emph{SmartProduct} backend, and the user's input which finalises the values of the XML document. 
The example shown in Figure~\ref{fig:epcis_aggregation_doc}, illustrates an \emph{Aggregation event} generated when a transporter loads various products into a truck, identified by a serial number, and a batch of products in the same pallet. The serial numbers of the products are "urn:epc:id:sgtin:123456.7123883.111" and "urn:epc:id:sgtin:123456.7123883.222", while the batch number is "urn:epc:class:lgtin:049111.9123456.7ABC", for which a quantity element of 30 items is also specified. In this XML document, the pallet unique serial number is denoted as parentID "urn:epc:id:sscc:103456.0123456789", and the serial number of the truck is "urn:epc:id:sgtin:401111.4444444.5V9K662R66". The status of the event is indicated by the \emph{bizStep} and the \emph{disposition}, which are set to "packing" and "in-progress", respectively. 


Figure~\ref{fig:application_overview} shows the procedure of tracking the storing of products in a warehouse using the mobile application; starting from top left to bottom right through six steps: Step 1: Overview of the homepage of a user, Step 2: The user selects to check-in products in the warehouse, Step 3: The user has to either scan the location of the warehouse (e.g. encoded in a QR code tag), or select one from a pre-populated list, Step 4: The user selects the tag type to scan (QR, BR, or NFC), Step 5: The list of the items scanned appears, and the proceed/save button is pressed, and Step 6: Communication with the \emph{SmartProduct} platform takes place and products' storing finalizes.

The availability of the backend or the general connectivity with the Internet, cannot be always guaranteed; however, EPCIS events' recording should not be detained in such cases. By leveraging Android’s diversity in storing data, the \emph{Shared Preferences}\footnote{\url{https://developer.android.com/reference/android/content/SharedPreferences}} feature is used, to store events locally in the device, while the network connection is lost, giving the ability to a user, to continue capturing events even offline. When the connection is restored, the events are securely sent to the \emph{EPCIS Capture Interface} of the platform, using the HTTPs protocol.

\subsection{Web frontend for authorised users}
\label{sec:interface)}
\begin{figure}
\includegraphics[scale = 0.25]{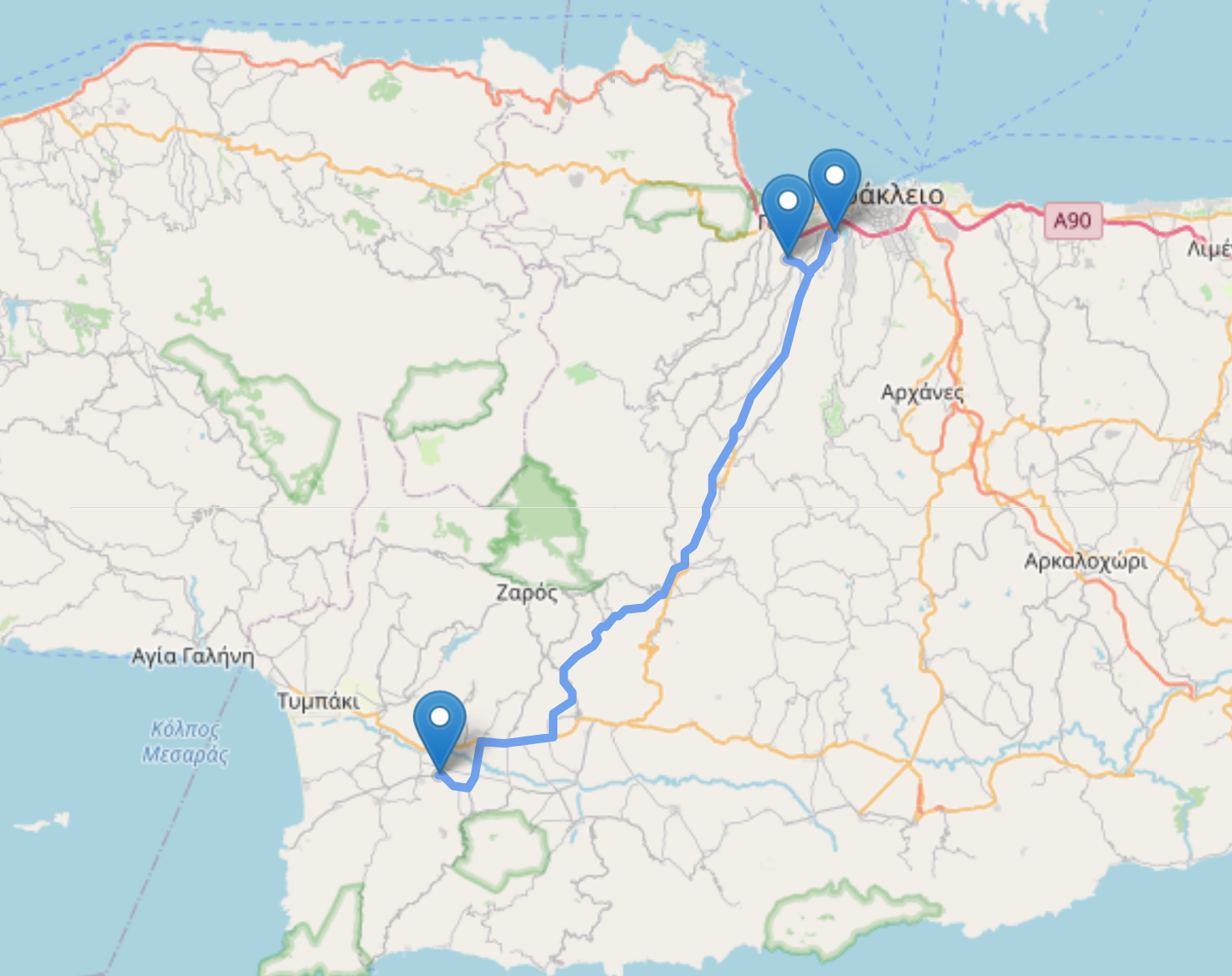}
\centering
\caption{Product tracking information displayed on map}
\label{fig:snap3}
\end{figure}

\begin{figure*}
\includegraphics[scale = 0.15]{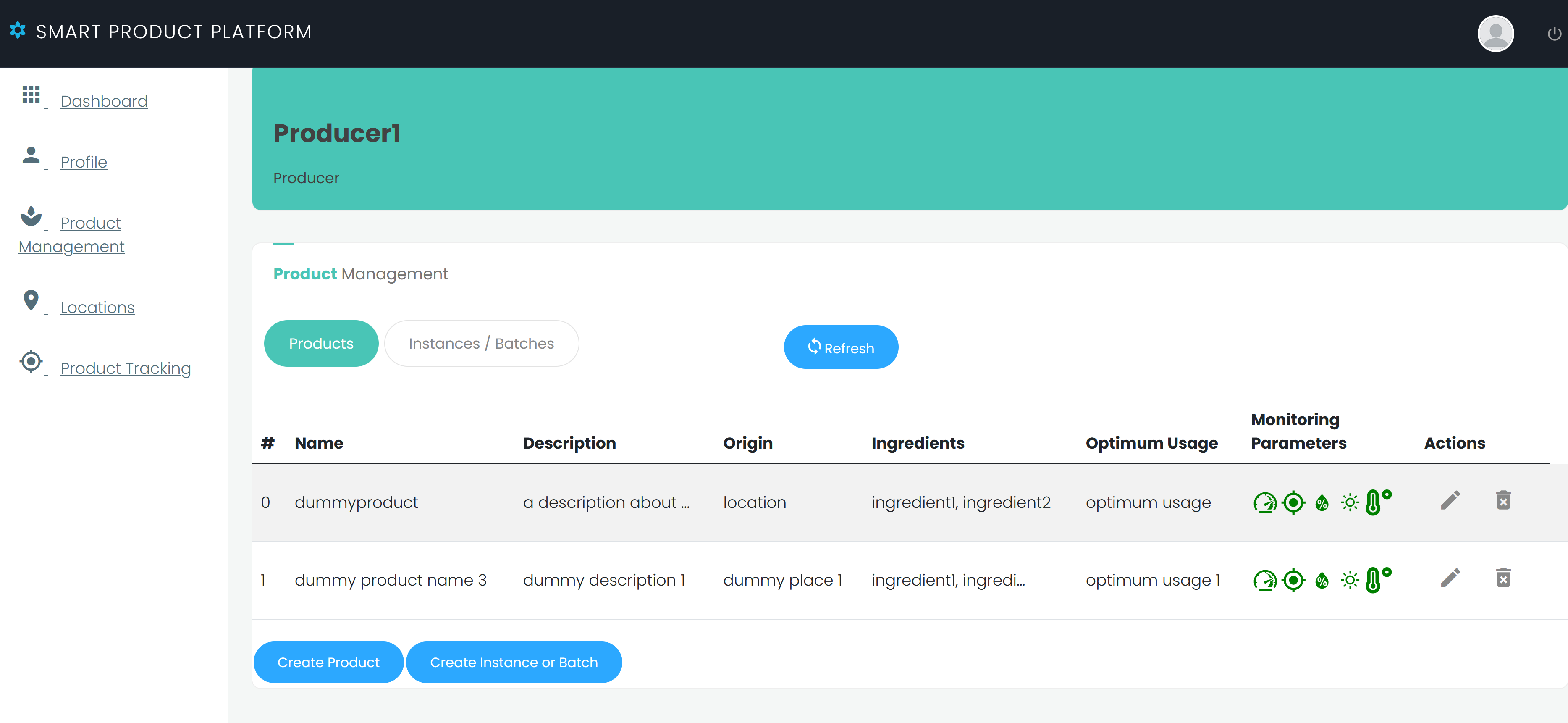}
\centering
\caption{Product creation}
\label{fig:snap1}
\end{figure*}

\begin{figure*}
\includegraphics[scale = 0.15]{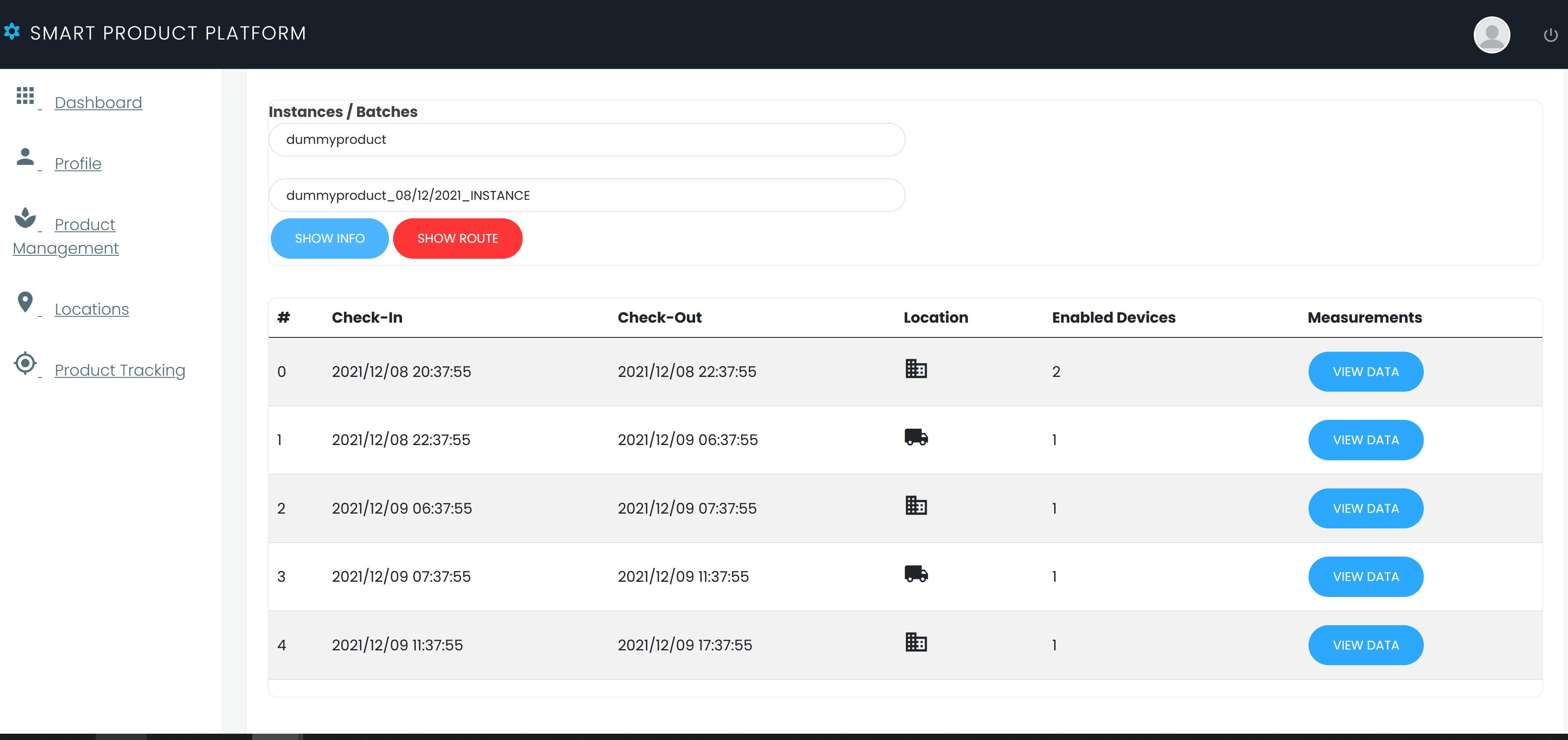}
\centering
\caption{Product tracking on different locations and vehicles}
\label{fig:snap2}
\end{figure*}

\begin{figure*}
\includegraphics[scale = 0.18]{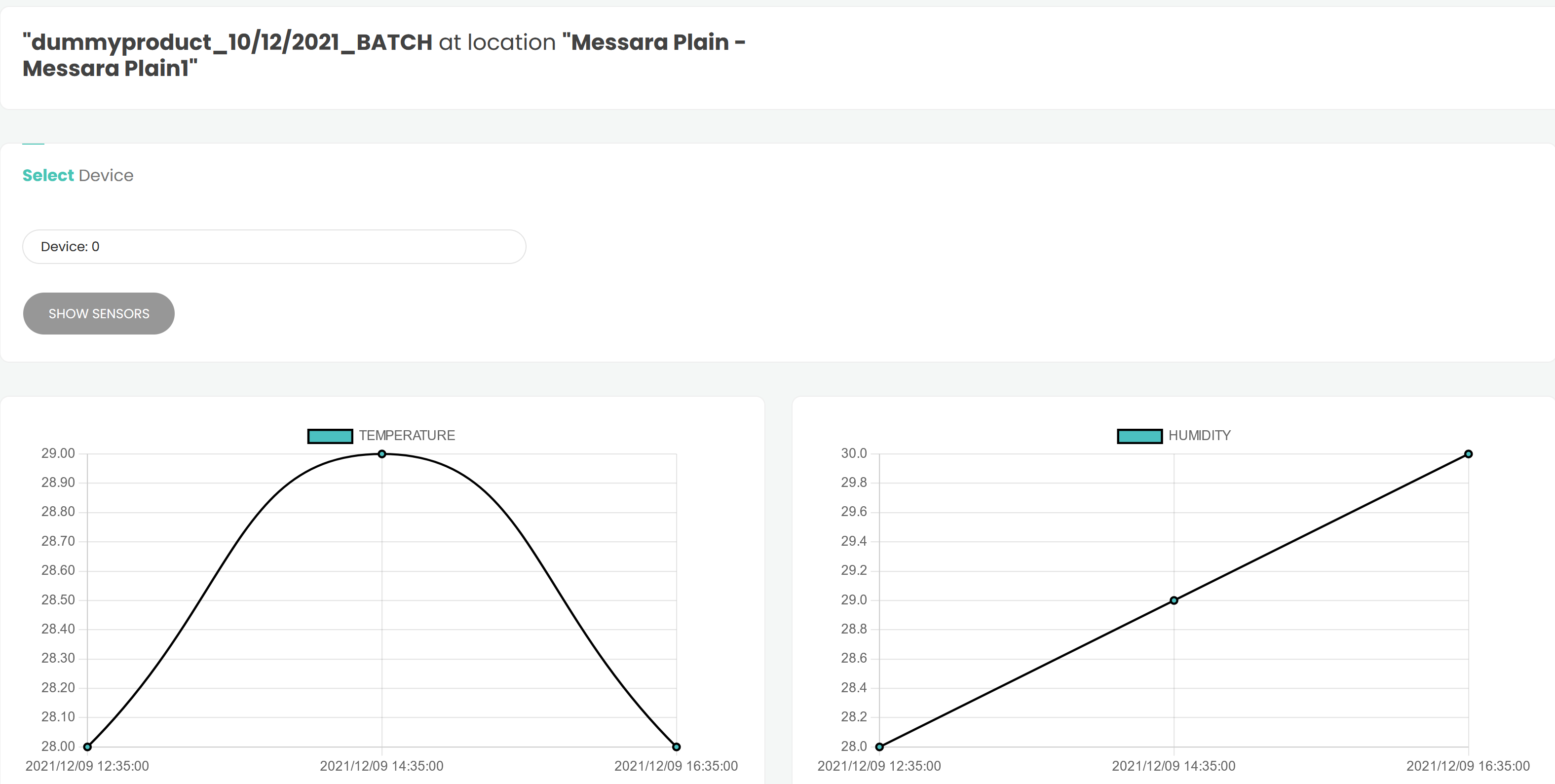}
\centering
\caption{IoT sensory data collected in a specific location where the product was stored}
\label{fig:snap4}
\end{figure*}

We have also developed a web frontend, based on a custom HTML template and React\footnote{\url{https://reactjs.org}}, which allows authenticated users to: (i) create (register) products, (ii) insert various useful details about them (description, origin, ingredients, optimum usage, etc.), (iii) create batches or instances for the created product, and (iv) select desired parameters to be monitored during product's journey within the supply chain (temperature, humidity, acceleration, geolocation, etc.). Figure~\ref{fig:snap1} shows a snapshot with several (example) products created. In Figure~\ref{fig:snap2}, we show the tracking of a specific instance of a product with details on check-in and checkout dates, type of locations (warehouse or vehicle), and the number of enabled IoT devices at each location. Moreover, we provide product tracking information displayed on a map (Figure~\ref{fig:snap3}), and IoT sensory data collected in a specific location (Figure~\ref{fig:snap4}).

\section{Conclusion and further work}
\label{sec:concl}
In this paper, we presented \emph{SmartProduct}, a platform for PMaaS, which leverages IoT technologies and the EPCIS standard. We provided a detailed analysis on platform's main components and services, the IoT device architecture, and the interactions of the various services. Moreover, two end-user interfaces were presented: an Android-based application to capture EPCIS events and thus to enable SCT, and a web frontend, which enables authenticated users to perform several operations on product registration and tracking. Further work includes various improvements and enhancements, such as the automation of IoT device registration, support for location registration based on GS1-compatible format, etc., and perform a thorough performance evaluation using suitable metrics.

\section*{Acknowledgments}
{This research has been partially financed by the European Union and Region of Epirus funds (project code: H$\Pi$1AB-0028183).}

\bibliographystyle{elsarticle-num}
\bibliography{bibliography.bib}

\end{document}